\documentclass[journal=jpcafh,manuscript=article,layout=twocolumn, 10pt]{achemso}

\usepackage{chemformula} 
\usepackage[T1]{fontenc} 

\usepackage{amsmath}
\usepackage{amsfonts}
\usepackage{amssymb}

\usepackage{graphicx}

\input{Qcircuit.tex}

\usepackage{multirow}

\author{Seyed Ehsan Ghasempouri}
\email{ehsan.ghasempouri@unb.ca}
\affiliation[unbchem]{Department of Chemistry, University of New Brunswick, 30 Dineen Dr Fredericton, New Brunswick E3B 5A3 Canada}
\author{Gerhard W. Dueck}
\email{gdueck@unb.ca}
\affiliation[unbcomp]{Faculty of Computer Science, University of New Brunswick, 550 Windsor St, Fredericton, New Brunswick E3B 5A3 Canada}
\author{Stijn De Baerdemacker}
\email{stijn.debaerdemacker@unb.ca}
\affiliation[unbchem]{Department of Chemistry, University of New Brunswick, 30 Dineen Dr Fredericton, New Brunswick E3B 5A3 Canada}
\alsoaffiliation[unbmath]{Department of Mathematics and Statistics, University of New Brunswick, P.O. Box 4400, Fredericton, New Brunswick E3B 5A3 Canada}

\title{Modular cluster circuits for the Variational Quantum Eigensolver}


\begin{document}

\begin{tocentry}

\begin{equation}
\begin{array}{c}
\Qcircuit @C=1em @R=.7em {
 &\gate{R_Y(\theta_0)} & \ctrl{1} & \gate{R_Y(\theta_4)} & \qw & \ctrl{3} & \qw \\
 &\gate{R_Y(\theta_1)} & \targ & \gate{R_Y(\theta_5)} & \ctrl{1} & \qw & \qw \\
 &\gate{R_Y(\theta_2)} & \ctrl{1}  & \gate{R_Y(\theta_6)} & \targ & \qw & \qw \\
 &\gate{R_Y(\theta_3)} & \targ & \gate{R_Y(\theta_7)} & \qw & \targ  & \qw \gategroup{2}{4}{3}{5}{0.7em}{--} 
}
\end{array}\notag
\end{equation}

\end{tocentry}

\begin{abstract}
The variational quantum eigensolver (VQE) algorithm recently became a popular method to compute quantum chemical properties of molecules on noisy intermediate scale quantum (NISQ) devices.   In order to avoid noise accumulation from the NISQ device in the quantum circuit, it is important to keep the so-called quantum depth of the circuit at a minimum, defined as the minimum number of quantum gates that need to be operated sequentially. In the present work, we introduce a modular 2-qubit cluster circuit that allows for the design of a shallow-depth quantum circuit compared to previously proposed architectures without loss of chemical accuracy.  Moreover, by virtue of the simplicity of the cluster circuit, it is possible to assign a valence bond chemical interpretation to the cluster circuit.  The design was tested on the \ch{H2}, \ch{(H2)2} and \ch{LiH} molecules, as well as the finite-size transverse-field Ising model, as the latter provides additional insights in the construction of the circuit in a resonating valence bond picture.  
\end{abstract}

\section{Introduction}
\label{sec:1}
The first idea of a quantum computer was introduced as early as the 1980s\cite{benioff:1980,manin:1980,feynman:1981} when it was argued that the ideal device on which to simulate a quantum system would be an \emph{other} quantum system.  These devices would harness the fundamental principles of superposition and entanglement that make simulating quantum systems exponentially hard in general on classical von Neumann devices.  Nevertheless, the past century has provided us with extraordinary progress in the development of classical quantum many-body approaches and electronic structure methods \cite{helgaker:2014} allowing us to simulate a majority of molecular systems on classical devices, however still at a polynomially scaling cost, and excluding the class of strongly correlated quantum systems.  The latter systems are characterized by a proliferation of quasi-degenerate electron configurations that quickly grow beyond the processing and memory capacities of classical devices.  Well known examples of such systems are transition metal compounds, important for photo luminescence \cite{biju:2008}, batteries \cite{wang:2021} or catalysis \cite{norskov:2009} applications.  For this reason, there still is strong interest in the development and investigation of new methods and technologies that have these strong quantum correlations baked in, in which quantum computing is a prime candidate. 

Over the past few decades, several physical realizations and architectures for a potential quantum computer have been proposed, cumulating in a recent surge in so-called noisy intermediate quantum devices (NISQ) \cite{preskill:2018,preskill:2023}.  These devices are characterized by quantum gates with finite accuracy or fidelity operating on a limited number of logical qubits, so applications are presently restricted to small molecules that can be simulated with compact quantum circuits.  The current method of choice on these devices is the variational quantum eigensolver (VQE)\cite{peruzzo:2014,mcclean:2016}, a quantum-classical hybrid approach to compute the energy of a molecular system.  In the VQE, a trial wave function is implemented by means of elementary quantum gates on a quantum device which is then iteratively optimized using classical optimization algorithms.  The optimization is performed variationally on the energy expectation value of the Hamiltonian which is evaluated indirectly through measurements of the density matrix on the quantum device.  The major benefit of implementing the wave function on a quantum computer, rather than on a classical device, is that the quantum device provides access to classes of quantum many-body wave functions that were unreachable by classical approaches in the pre-NISQ era.  

A good example is the unitary coupled cluster ansatz (UCC) \cite{bartlett:1989,bartlett:2007,seeley:2012,harsha:2018,barkoutsos:2018}, a variation of the highly successful coupled cluster method in which the particle-hole operators are put in anti-Hermitian shape.  Interestingly, the unitary properties of the exponentiated particle-hole operators make them perfectly suited as unitary quantum gates acting on the reference state.  The UCC at the singles and doubles level (UCCSD)\cite{romero:2018,ryabinkin:2018,lang:2020} enjoys the benefit of its close connection to the traditional coupled cluster singles and doubles (CCSD) method\cite{bartlett:2007}, however it suffers from an excessive number of quantum gates and elevated circuit complexity if all particle-hole excitations are to be considered.  Consequently, the UCCSD ansatz has a high circuit depth, which is defined as the minimum number of serial quantum gates within the circuit.  On NISQ devices, the circuit depth needs to be kept as low as possible in order to mitigate the build up of noise.  One approach to minimize the number of particle-hole operators is the use of a greedy optimization algorithm such as the ADAPT-VQE \cite{grimsley:2019,tang:2021,anastasiou:2022}, in which the circuit of concatenated particle-hole operators is organically grown by means of intermediate evaluations of the energy gradient at each step of the circuit design\cite{fedorov:2022}.  The elegance of the ADAPT-VQE approach is contained in the efficient evaluation of the energy gradient in terms of measurements.  Unfortunately, symmetries are not optimally encoded in the algorithm, and the greedy aspect of the optimization procedure does not guarantee convergence to the global minimum \cite{burton:2022}.  A closely related approach is the iterative qubit coupled cluster (iQCC) method\cite{ryabinkin:2020,lang:2020}, in which the quantum depth or scale of the quantum circuit is kept fixed, however the complexity of the wave function is encapsulated in a similarity transformation of the Hamiltonian.  As a result, despite having a significantly lower circuit depth, additional measurements are required as the dressed Hamiltonian grows in terms of Pauli strings \cite{zhang:2022}.  In addition, the reduction in circuit depth does not automatically lead to improved convergence with respect to the ADAPT-VQE \cite{ryabinkin:2020,zhang:2022}.  

In the present work, we introduce a minimal cluster circuit for the VQE that will allow for a compact and modular circuit design with reduced circuit depth.  The minimal design of the circuit allows for a chemical interpretation in terms of valence bond (VB) structures \cite{cooper:2002,wu:2011}.  The textbook picture of a VB structure is a resonance between different electron pair configurations and provides a chemically inspired means to construct compact wave functions.  Paired electron wave function ans\"atze have been considered, both in the context of unitary pair coupled cluster doubles (UpCCD) wave functions \cite{lee:2019} as well as antisymmetrized geminal powers (AGP)\cite{khamoshi:2021,khamoshi:2023}, however with a circuit complexity that is comparable to UCC methods or with circuit designs that explicitly target certain electron pairing and valence bond resonance schemes such as in the separable pair approximations (SPA) \cite{kottmann:2022a} and graph-based approach \cite{kottmann:2022b}.  We will make a connection with the resonating valence bond (RVB) method\cite{anderson:1973}, applied to the finite-size transverse-field Ising model, known from condensed matter physics and statistical physics\cite{pfeuty:1970}.   

The use of cluster circuits has been proposed before in the ClusterVQE method \cite{zhang:2022}, however from a top-down point of view in which larger spaces are split into simpler building blocks using mutual information indices.  Mutual information indices are insightful as \textit{a posteriori} descriptors of strong correlation in classical molecular wave function methods, most particularly in the context of the density matrix renormalization group method (DMRG)\cite{barcza:2011,boguslawski:2012}.  However, their application in the \textit{a priori} identification of computational wave function simplifications has limitations, as the optimal entanglement scheme can often only be identified after a more sophisticated computation has been performed.  The simplification in the ClusterVQE method is achieved by means of a decomposition in fully parallel cluster circuits with shorter circuit depth, however at the cost of a renormalization of the Hamiltonian, leading to an increase in number of Pauli words to be measured.  As such the ClusterVQE can be seen as a trade off between the iQCC and conventional UCC approaches in terms of circuit depth versus number of Pauli words.  In the present work, we aim to keep the circuit depth low while refraining from any Hamiltonian renormalization, keeping the number of Pauli words equal to, e.g., the ADAPT-VQE approach.  For this, we employ a bottom-up approach in terms of modular clusters.  As can be expected, the choice of modular clusters inserts a bias into the design of the circuit, which can be mitigated by means of a greedy optimization procedure like the ADAPT-VQE.  However, greedy optimization procedures are tailored towards individual Hamiltonians, and are unlikely to produce consistent results over a full range of Hamiltonian parameters, for instance for all possible bond stretches \cite{burton:2022}.  For this reason, our circuit design will be informed by chemical input from VB structures, however allowing for sufficient flexibility such that it is potentially applicable among different chemical systems and configurations.  

It is worth noting that other measures besides the circuit depth exist to reduce the complexity of quantum circuits on NISQ devices.  For instance, one can also target the circuit width, which refers to the number of physical qubits on the real quantum device. Several efforts have been made to use symmetries to reduce the number of qubits \cite{bravyi:2017,yen:2019,setia:2020,fujii:2022,zhang:2021}. However, these methods are capable of reducing just a few number of qubits.  Recent advancements allowed for further reduction of the circuit width by concatenating different VQE approaches to solve an effective Hamiltonian, using tensor-network methods, or by reverting back to 1st quantization approaches \cite{liu:2019,barratt:2021,yuan:2021,eddins:2022,delgado:2022}.  In this work, we will let the circuit width be dictated by the fermion-qubit mapping, which is preferably local, in order to retain a close connection to the underlying orbital scheme.  

The rest of the paper is organised as follows.  We will define the cluster circuit in the following section in line with the exact quantum circuit for the 2-site transverse-field Ising model and the \ch{H2} molecule in a minimal basis set, and give it an interpretation in terms of valence bond resonances. In the next section, we employ the cluster circuit as a modular block to design full cluster circuits for multiple-qubit systems based on.  In the final section, we comment on the potential and limitations of the cluster circuits.   All simulations have been performed in the Qiskit environment\cite{qiskit:2023}, and can be accessed in the supplemental information.

\section{Cluster Circuits}\label{sec:2}

\subsection{Valence Bonding}

The quintessential minimal model of the \ch{H2} molecule is a superposition of a pair of electrons in the $\sigma$ bonding and $\sigma^\ast$ antibonding orbital of the system
\begin{equation}
|\ch{H2}\rangle=\cos\tfrac{\theta}{2}|\sigma\bar{\sigma}\rangle +\sin\tfrac{\theta}{2}|\sigma^\ast\bar{\sigma}^\ast\rangle,\label{vb:H2}
\end{equation}
in which the ($i,\bar{i}$) notation is used to denote the two spin-up and spin-down partners respectively in the spatial orbital $i$.  In a minimal model, such as STO-$n$G, the bonding and antibonding orbitals are constructed from the single atomic orbitals on the individual hydrogens respectively.  In all our simulations, we employed the restricted Hartree-Fock (RHF) orbitals as the initial state orbitals in the circuit.  Representing state (\ref{vb:H2}) on a quantum computer depends on the mapping.  The most economical mapping in terms of circuit width and depth exhausts all possible symmetries and considers the states $|\sigma\bar{\sigma}\rangle$ and $|\sigma^\ast\bar{\sigma}^\ast\rangle$ as the $|0\rangle$ and $|1\rangle$ projections of a single qubit.  As such, the state (\ref{vb:H2}) can be represented as a simple 1-qubit gate on the $|0\rangle$ reference state
\begin{equation}
\begin{array}{c}
\Qcircuit @C=1em @R=.7em {
\lstick{\ket{0}} & \gate{R_Y(\theta)} & \qw  
}
\end{array},
\end{equation}
with $R_Y(\theta)$ a rotation around the $y$-axis 
\begin{equation}
R_Y(\theta)=\left(\begin{array}{cc}\cos\tfrac{\theta}{2} & -\sin\tfrac{\theta}{2} \\ \sin\tfrac{\theta}{2} & \cos\tfrac{\theta}{2}\end{array}\right).
\end{equation} 
This representation assumes a perfect pairing organization, in which each orbital will always be doubly occupied or empty.  However, the perfect pairing scheme is only exact in 2-electron systems, so multiple electron system require individual orbital mappings such as Jordan-Wigner (JW), parity transformation (PT) or Bravyi-Kitaev (BK) mappings\cite{bravyi:2002,tranter:2018}.  In order to keep a close connection to the constituent orbitals, we will restrict ourselves to the JW and PT mappings in this work.  For easy reference, the connections are listed in Table \ref{table:vb:mappings}.
\begin{table}[!htb]
\centering
\caption{Relation between the Jordan-Wigner, parity transformation and physical states for \ch{H2} in a minimal basis set}.\label{table:vb:mappings}
\begin{tabular}[t]{|l|cccc|cc|}
  \hline
   states & \multicolumn{4}{|c|}{JW} & \multicolumn{2}{|c|}{PT}   \\
  \hline\hline
  $|\sigma\bar{\sigma}\rangle$ & 0 & 1 & 0 & 1 & 0 & 1 \\
  $|\sigma^\ast\bar{\sigma}\rangle$ & 0 & 1 & 1 & 0 & 0 & 0 \\    
  $|\sigma\bar{\sigma}^\ast\rangle$ & 1 & 0 & 0 & 1 & 1 & 1 \\  
  $|\sigma^\ast\bar{\sigma}^\ast\rangle$ & 1 & 0 & 1 & 0 & 1 & 0 \\
  \hline
   & $\bar{\sigma}^\ast$ & $\bar{\sigma}$ & $\sigma^\ast$ & $\sigma$ & & \\
  \hline
\end{tabular}
\end{table}
As can be inferred from the Table, JW is a faithful 4-qubit representation of the occupancies of each orbital, whereas the 2-qubit PT mapping only keeps track of the occupancy of the $\sigma$ and $\bar{\sigma}^\ast$ orbital, ensuring that spin-projection symmetry automatically fixes the occupancy of the other two orbitals.   Indeed, the circuits based on the JW  mapping can break all symmetries related to the minimal \ch{H2} model, including particle number symmetry, whereas the PT conserves particle number and spin-projection symmetry, however not total spin symmetry.  Leaving the JW mapping to a following section, it is straightforward to check that the state (\ref{vb:H2}) can be obtained in the PT mapping as 

\begin{equation}
\begin{array}{c}
\Qcircuit @C=1em @R=.7em {
\lstick{\ket{0}}  &\gate{R_Y(\theta_0)} & \ctrl{1} & \qw\\
\lstick{\ket{0}}  &\gate{R_Y(\theta_1)} & \targ    & \qw 
}
\end{array}\label{vb:cluster}
\end{equation}
with $\theta_0=\theta$ and $\theta_1=\pi$ the parameters of the rotation gates, followed by the \texttt{CNOT} gate.  Therefore, the circuit (\ref{vb:cluster}) provides an exact representation of the bonding/antibonding structure of the \ch{H2} system.  We refer to this circuit as a cluster circuit unit, which will be used in the following sections as a modular form for other systems. 

\subsection{Transverse-Field Ising model}
Before moving to those other systems, we note that the circuit (\ref{vb:cluster}) also simulates the eigenstate of the 2-site transverse-field Ising model
\begin{equation}
\hat{H}_{\textrm{2-site IM}}= -J Z_0Z_1 - h(X_0+X_1)\label{tvIM:2siteHamiltonian},
\end{equation}
with $(X_i,Y_i,Z_i)$ the Pauli spin matrices on lattice site $i$, $J$ the nearest-neighbor magnetic interaction along the $z$-direction, and $h$ the magnetic field in the $x$-direction.  We will consider only antiferromagnetic interactions $J<0$.  The general form of the transverse-field Ising Hamiltonian for $n$ lattice sites is 
\begin{equation} 
 \hat{H}_{n\textrm{-site IM}} = -\sum_{i,j=1}^nJ_{ij}Z_iZ_j-h\sum_{i=1}^nX_i
\end{equation}
in which the interaction matrix $J_{ij}$ follows the topology of the lattice and $h$ is again the magnetic field in the $x$-direction.  The Hamiltonian describes a competition between the interaction terms which aim to (anti)align the $z$-projection of the lattice spins with one an other, and the magnetic field aligning all spins along the same $x$-direction.  The quantum transverse-field Ising model is a very simple and intuitive model in statistical physics and condensed matter physics that has been fertile for method development purposes, for instance the resonating valence bond method in condensed matter\cite{anderson:1973}.  We will argue that the finite-size transverse-field Ising model can provide similar insights for molecular systems

\begin{figure}[!htb]
\centering
\includegraphics{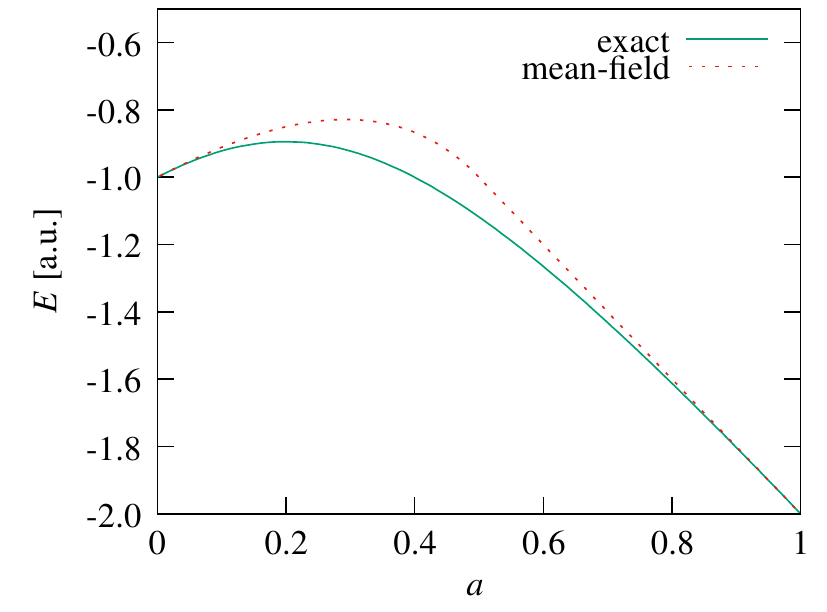}
\caption{\label{fig1} Exact energy (\ref{tvIM:energy}) and mean-field prediction (\ref{tvIM:meanfield}) of the 2-site transverse-field Ising model (\ref{tvIM:2siteHamiltonian}) as a function of $a\in[0,1]$ with $h=a$ and $J=a-1$.  The cluster circuit energy matches the exact energy.}
\end{figure}

The ground state of the 2-site model is given by
\begin{align}
|\psi_\textrm{2-site IM}\rangle=&\cos\theta\tfrac{1}{\sqrt{2}}(|00\rangle+|11\rangle)\notag\\
&+\sin\theta \tfrac{1}{\sqrt{2}}(|01\rangle+|10\rangle)\label{tvIM:2site-gs-state}
\end{align}
with $\tan\theta=\frac{J-\sqrt{J^2+4h^2}}{2h}$, and the associated energy given by
\begin{equation}
E_{\textrm{2-site IM}}=-\sqrt{J^2+4h^2}.\label{tvIM:energy}
\end{equation} 
The ground-state energy is plotted in Figure \ref{fig1}
as a function of the dimensionless parameter $a\in[0,1]$ with $h=a$ and $J=a-1$.  It is straightforward to check that state (\ref{tvIM:2site-gs-state}) can be obtained again by the cluster circuit (\ref{vb:cluster}) with $\theta_0=\frac{\pi}{2}$ and $\theta_1=2\theta$, showing that the cluster circuit is not only adequate in capturing the correlation in the minimal \ch{H2} molecule, but also the 2-site Ising model.  The superposition in the latter model however is between the anti-ferromagnetic spin configurations $\frac{1}{\sqrt{2}}(|01\rangle+|10\rangle)$ and the ferromagnetic $\frac{1}{\sqrt{2}}(|00\rangle+|11\rangle)$ counterparts. 

It is instructive to investigate the role of the \texttt{CNOT} gate in the cluster circuit (\ref{vb:cluster}), also referred to as the entangler gate.  For this purpose, we absorb the \texttt{CNOT} in the Hamiltonian via a similarity transformation
\begin{align}
\hat{H}^\prime_{\textrm{2-site IM}} &= \texttt{CNOT}^{-1}\hat{H}_{\textrm{2-site IM}}\texttt{CNOT}\\
&=-JZ_1-hX_1(1+X_0)\label{tfim:transformedH}
\end{align}
Interestingly, the transformed Hamiltonian (\ref{tfim:transformedH}) has again a transverse field Ising-like structure, however with the role of the magnetic field $h$ and magnetic interaction $J$ exchanged.  Moreover, the Hamiltonian supports a factorizable ground state 
\begin{equation}
|\psi^\prime_\textrm{2-site IM}\rangle=\tfrac{1}{\sqrt{2}}(|0\rangle+|1\rangle)(\cos\theta|0\rangle+\sin\theta|1\rangle)\label{tfim:mf}
\end{equation}
with again $\tan\theta=\frac{J-\sqrt{J^2+4h^2}}{2h}$, which is associated with a mean-field wave function. The state (\ref{tfim:mf}) can be represented in circuit form by
\begin{equation}
\begin{array}{c}
\Qcircuit @C=1em @R=.7em {
\lstick{\ket{0}} & \gate{R_Y(\theta_0)} & \qw\\
\lstick{\ket{0}} & \gate{R_Y(\theta_1)} & \qw 
}
\end{array}\label{vb:mean field}
\end{equation}
with the same $\theta_0=\frac{\pi}{2}$ and $\theta_1=2\theta$.  We notice that incorporating the \texttt{CNOT} entangler in the Hamiltonian reduces the circuit depth from 2 to 1 without increasing the complexity of the Hamiltonian in terms of Pauli words or strings.  Unfortunately, this is a feature that is unique to a 2-qubit Hamiltonian as the Hamiltonian complexity increases in general, such as is the case in the iQCC\cite{lang:2020,ryabinkin:2020}. 

The role of the entangler in the circuit design becomes clearer when employing the circuit (\ref{vb:mean field}) directly to the Hamiltonian.  This is exactly the mean field approximation to the Hamiltonian, with energy expectation value \\
\begin{equation}
E[\theta_0,\theta_1]=-J\cos\theta_0\cos\theta_1-h(\sin\theta_0+\sin\theta_1),
\end{equation}
leading to the solutions
\begin{equation}
\left\{\begin{array}{ll}\theta_0+\theta_1=\pi\ \textrm{and}\ \sin\theta_0=-\frac{h}{J}, & \forall (-\frac{h}{J})<1, \\
                        \theta_0=\theta_1=\frac{\pi}{2}, &\forall (-\frac{h}{J})\ge 1,\end{array}\right.
\end{equation}
for the two domains separated by the $(-\frac{h}{J})=1$ (or $a=\frac{1}{2}$) point.  For $(-\frac{h}{J})<0$ the magnetic spin-spin interaction is dominant, whereas the magnetic field dominates in the $(-\frac{h}{J})>0$ regime (see further). 
The energy solutions for both solutions are 
\begin{equation}
E_{\textrm{MF}}=\left\{\begin{array}{ll} \frac{h^2}{J}+J & \forall (-\frac{h}{J})<1 \\
                        -2h &\forall (-\frac{h}{J})\ge 1\end{array}\right.,\label{tvIM:meanfield}
\end{equation}
which are also plotted in Figure \ref{fig1}.  The difference between the exact energy and the mean field approximation is referred to as the correlation energy  
\begin{equation}
E_c=|E_{\textrm{exact}}-E_{\textrm{MF}}|.
\end{equation}
It can be seen from Figure \ref{fig1} that the exact energy is well reproduced in the $a\rightarrow 0$ ($h\rightarrow0$) and $a\rightarrow1$ ($J\rightarrow0$) regime.  However, the correlation energy is largest in the intermediate regime where correlation is largest, emphasizing the importance of the entangler \texttt{CNOT} in getting the correct resonance structure. 

\begin{figure}[!htb]
\centering
\includegraphics{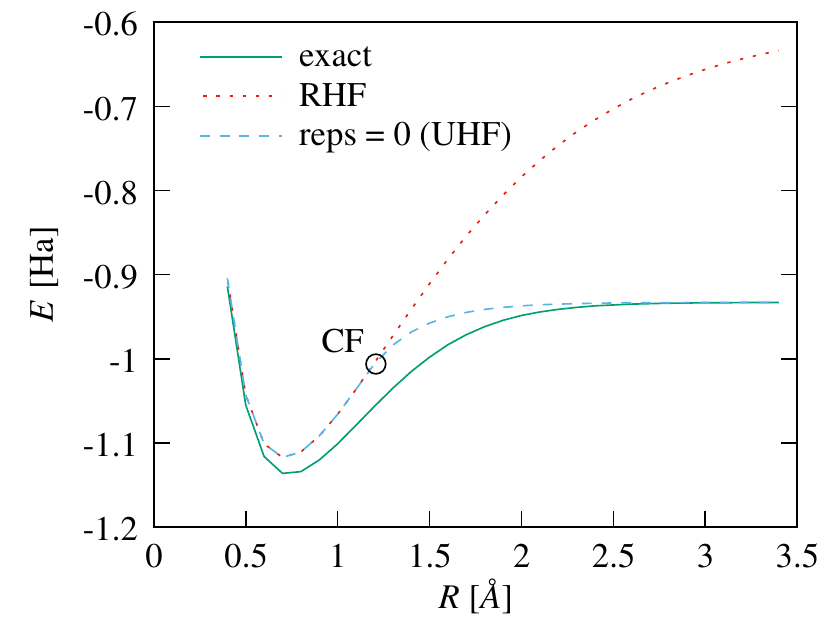}
\caption{\label{fig2} Energy profile as a function of bond distance of the \ch{H2} molecule in the STO-3G basis for the exact (full line), $\textrm{reps}=0$ (dashed line) and initial state (RHF, dotted line).  The $\textrm{reps}=0$ (mean field) and initial state curves are equivalent to the UHF and RHF respectively.  The Coulson-Fischer point is marked with CF.  The cluster circuit (\ref{vb:cluster}) reproduces the exact result by construction.}
\end{figure}

From the mean field solution, we observe a symmetry breaking process around the $-\frac{h}{J}=1$ point, as both spins have the same orientation $\theta_0=\theta_1=\frac{\pi}{2}$ along the $x$-axis for $-\frac{h}{J}>1$, whereas the symmetry is spontaneously broken in the other regime.  This phenomenon is very well understood in quantum chemistry in the context of spin-restricted (RHF) and unrestricted (UHF) Hartree-Fock theory \cite{coulson:1949}.  Indeed, applying the mean-field circuit (\ref{vb:mean field}) to the \ch{H2} Hamiltonian in the PT mapping produces exactly the UHF state, as the rotation gates $R_Y(\theta_i)$ (with $\theta_0\neq\theta_1$) produce different linear combination between the bonding and antibonding orbitals for the spin-up and spin-down electrons.  The mean-field state (\ref{vb:mean field}) in the PT mapping for \ch{H2} (see Table \ref{table:vb:mappings}) becomes
\begin{align}
|\ch{H2}_{(\textrm{UHF})}\rangle&=\hat{R}_{Y}(\theta_0)\hat{R}_{Y}(\theta_1)|00\rangle\\
&=(\cos\theta_0|\sigma^\ast\rangle+\sin\theta_0|\sigma\rangle)\\
&\qquad\times(\cos\theta_1|\bar{\sigma}\rangle+\sin\theta_1|\bar{\sigma}^\ast\rangle),
\end{align}
allowing for an asymmetric localization of the orbitals for the spin-up and -down electrons.  For instance, the $\theta_0=-\theta_1=\frac{\pi}{2}$ solution leads to a complete localization of the electrons on the opposite hydrogens in the molecule.  It is well known that UHF does not provide a variational advantage with respect to RHF around the equilibrium distance ($R_{\textrm{eq}}=0.718\textrm{\AA}$), however at the so-called Coulson-Fischer (CF) point ($R_{\textrm{CF}}=1.191\textrm{\AA}$), the spin symmetry is spontaneously broken at the mean field level, giving rise to an energetic preference of UHF over RHF (see Figure \ref{fig2}) \cite{lykos:1963}.  It is interesting coincidence that the RHF energy is reproduced by evaluating the energy directly from the $|\sigma\bar{\sigma}\rangle=|01\rangle$ state in the PT mapping (see Table \ref{table:vb:mappings}), or equivalently, $\theta_0=\pi$ and $\theta_1=0$ in circuit (\ref{vb:mean field}).
In conclusion, the interpretation of the cluster circuit as a concatenation of single-qubit rotation gates with a \texttt{CNOT} entangler is in line with a VB picture\cite{cooper:2002}, as the former induce a localisation of the bonding/antibonding orbitals engaging in the bond, and the latter produces the actual resonance and correlation. 

This basic insight in the cluster circuit will provide the rationale behind the circuit designs for multi-qubit systems that will be discussed in the following section.

\section{Modular Circuits}

The question is now how the framework from the previous section can be extended to other molecular systems.  We can find inspiration from the $n$-site Ising model, with $n$ the number of qubits needed in the circuit.  

\subsection{4-qubit systems}
\begin{figure}[h]
\begin{center}
  \hfill
    \includegraphics[width=0.3\linewidth]{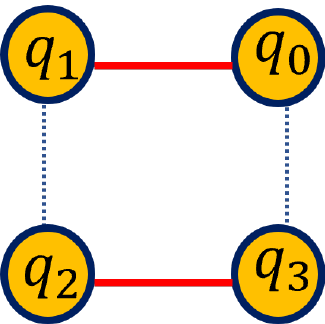}
  \hfill
    \includegraphics[width=0.3\linewidth]{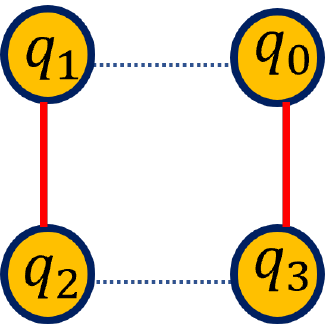}
  \hfill 
  
  \end{center}
  \caption{\label{fig3-4sitervb}RVB structures of the 4-site ($q_i,i=1\dots4)$ transverse-field Ising model.  Each pair of qubits connected with a red full line represent a cluster circuit unit (\ref{vb:cluster}).}
\end{figure}

To fix ideas, one can design a circuit for the \ch{H2} in the STO-3G basis using the JW mapping instead of the PT mapping, requiring four qubits instead of two (see Table \ref{table:vb:mappings}).  For this purpose, we focus on the 4-site transverse field Ising model with periodic boundary conditions first, more in particular on the RVB picture of the model (see Figure \ref{fig3-4sitervb}, in which the lattice sites are denoted by $q_i$ ($i=0\dots3$).  
The RVB picture of this system consists of a resonance between two 2-site bonding configurations which distinguish themselves by the choice of neighbour with whom they interact.  This suggests the combination of two blocks of two cluster circuits (\ref{vb:cluster}), as in the following circuit
 \scriptsize
\begin{equation}
\begin{array}{c}
\Qcircuit @C=1em @R=.7em {
\lstick{\ket{0}}  &\gate{R_Y(\theta_0)} & \ctrl{1} &\qw & \gate{R_Y(\theta_4)} & \qw & \ctrl{3} & \qw & \gate{R_Y(\theta_8)}& \qw\\
\lstick{\ket{0}}  &\gate{R_Y(\theta_1)} & \targ &\qw & \gate{R_Y(\theta_5)} & \ctrl{1} & \qw & \qw& \gate{R_Y(\theta_{9})}& \qw\\
\lstick{\ket{0}}  &\gate{R_Y(\theta_2)} & \ctrl{1} &\qw & \gate{R_Y(\theta_6)} & \targ & \qw & \qw& \gate{R_Y(\theta_{10})}& \qw\\
\lstick{\ket{0}}  &\gate{R_Y(\theta_3)} & \targ &\qw & \gate{R_Y(\theta_7)} & \qw & \targ  & \qw& \gate{R_Y(\theta_{11})}& \qw\gategroup{1}{2}{2}{3}{0.7em}{--} \gategroup{3}{2}{4}{3}{0.7em}{--}
}
\end{array}.\label{vb:4site-ising}
\end{equation}
\normalsize
The full circuit consists of two layers of cluster circuits with an additional layer of single-qubit rotation gates.  Each layer represents a RVB configuration from Figure \ref{fig3-4sitervb}, consisting of cluster circuit units (\ref{vb:cluster}) that represent each of the two connections.  For instance, the cluster circuit units corresponding to the left configuration in Figure \ref{fig3-4sitervb} are highlighted by dashed boxes.   The first layer's \texttt{CNOT} entangler map differs from the second layer.  A pair of interactions exist in the first layer between qubits $q_0$ and $q_1$, and an other pair between $q_2$ and $q_3$, whereas the second layer connects $q_0$ with $q_3$ and $q_1$ with $q_2$.  The benefit of this VB driven design is that the cluster circuits are implemented in parallel, therefore limiting the circuit depth per layer to just two.  Therefore, it is possible to concatenate multiple layers, increasing the variational degrees of freedom without proliferating the number of entanglers and circuit depth.  This is a considerable advantage of the modular cluster circuit design over, for instance, unitary coupled cluster quantum circuits.  The total number of layers will be referred to as the repetitions (reps) of the circuit.  The inclusion of a final layer of rotation gates allows one to associate the $\textrm{reps}=0$ circuit with the mean-field solution. 

\begin{figure}[!htb]
\centering
\includegraphics{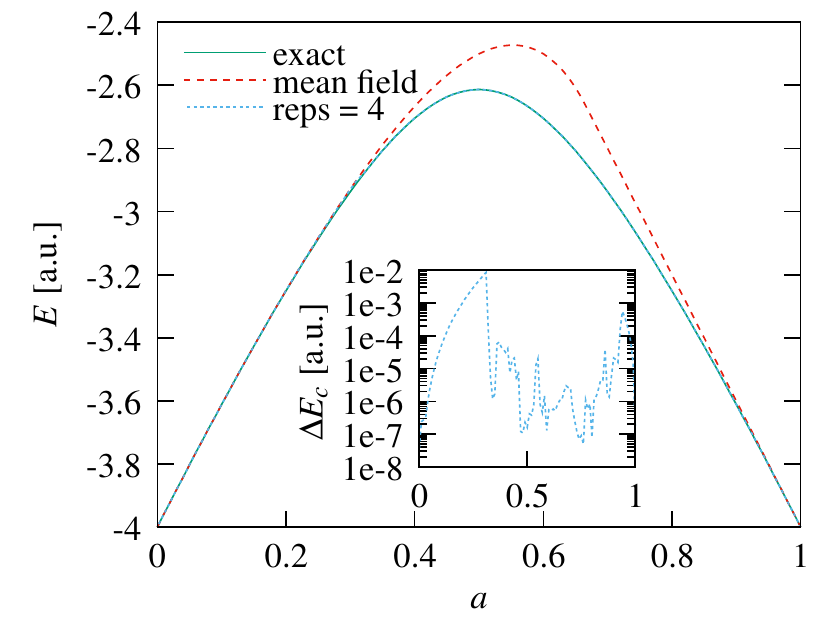}
\caption{\label{fig4-4siteising}Energy predictions for the 4-site transverse-field Ising model with the full cluster state with $\textrm{reps}=4$ (blue dotted line) and mean-field circuit with $\textrm{reps}=0$ (red dashed line), confronted with the exact energy (green full line).  The missing correlation energy (\ref{4qubit:deltaE}) $\textrm{reps}=4$ is given in the inset.}
\end{figure}

The 4-site transverse Ising model was simulated using the circuit (\ref{vb:4site-ising}) in which layer one and two were each repeated once ($\textrm{reps}=4$).  The result of the full cluster circuit is confronted with the exact energy in Figure \ref{fig4-4siteising},
together with the mean-field predictions, consisting of a single layer of single-qubit gates without entanglers ($\textrm{reps}=0$).  It can be seen from the figure that the full cluster circuit reproduces the exact energy up to $0.01$a.u.\ across all values of the parameter $a$.  For easy evaluation, we also define and provide the missing correlation energy 
\begin{equation}
\Delta E_c = |E_{\textrm{exact}}-E_{\textrm{VQE}}|\label{4qubit:deltaE}
\end{equation}
in all the plots throughout this paper for all presented reps.  The circuit (\ref{vb:4site-ising}) has been explored in the context of the axial next-nearest neighbour Ising (ANNNI) model\cite{copetudoespinosa:2019}, however with \texttt{CZ} gates instead of \texttt{CNOT} (or \texttt{CX}). 

All numerical simulation were run in QASM simulator in the Qiskit \cite{qiskit:2023} environment.  For the VQE optimization, the sequential least squares programming (SLSQP) \cite{kraft:1988} library was used as a classical optimizer with a maximum number of 200 iterations.  We employed two different strategies for the initial guess for the rotation gate angles.  For the first (default) strategy, the initial rotation angles of the chosen ansatz for the VQE calculations are generated randomly between $[0,2\pi]$ for each value of the model parameter.  The second (adiabatic) strategy carefully selects a point in the model parameter space for which the method is likely to produce accurate results using the default strategy, for instance the equilibrium bond distance $R_{\textrm{eq}}$ for a diatomic molecule, and adiabatically changes the parameter to the desired value.  The advantage of the default method is that any point in the parameter space can be targeted in a one-shot calculation, whereas the benefit of the adiabatic approach is an enhanced resilience against local minima in the optimization process.  We will use the default optimization strategy in all following simulations, unless local minima become overly pronounced, in which we will also employ the adiabatic approach.

\begin{figure}[!htb]
\centering
\includegraphics{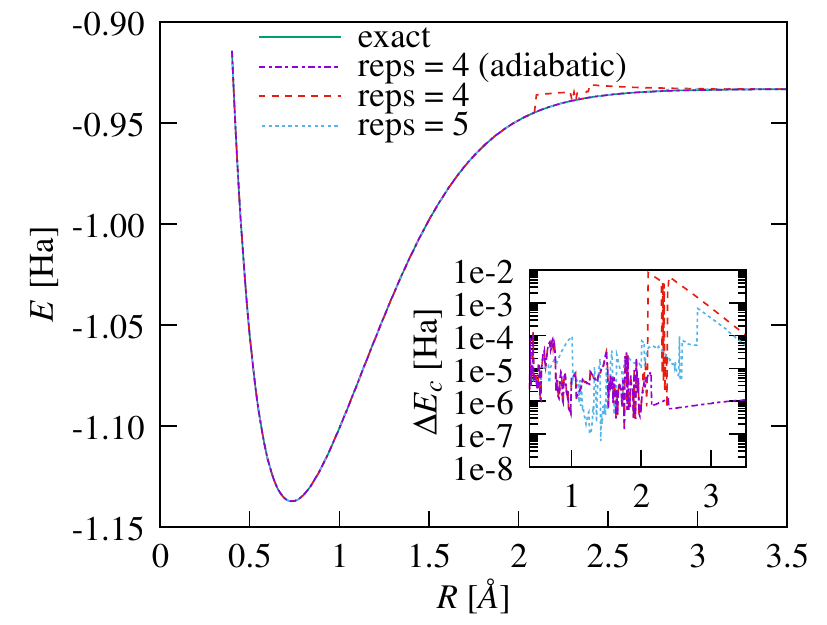}
\caption{\label{fig5-h2-sto3g-jw}Energy prediction of the full cluster circuit for the \ch{H2} in STO-3G basis using JW mapping for $\textrm{reps}=4$ (red dashed line), $\textrm{reps}=5$ (blue dotted line) and $\textrm{reps}=4$ (adiabatic) (purple dot-dashed line).  The exact reference energy is given by the green full line.  The inset shows the missing correlation energy (\ref{4qubit:deltaE}) for both reps in atomic units.  The $\textrm{reps}=5$ and $\textrm{reps}=4$ (adiabatic) energy predictions are within chemical accuracy ($\sim1\textrm{mHa}$).}
\end{figure}

In the spirit of the previous section, in which we recognized the universal nature of the cluster circuit for a molecular and magnetic 2-qubit system, we can employ the constructed 4-qubit circuit from the 4-site transverse field Ising model to target molecular 4-qubit systems.  As already mentioned earlier, the \ch{H2} molecule in the minimal STO-3G basis set with the JW mapping is a 4-qubit system.  As can be observed in Figure \ref{fig5-h2-sto3g-jw},  
the same modular cluster circuit architecture performs equally accurate at the $\textrm{reps}=4$ level, over the majority of bond lengths, and requires only one additional repetition ($\textrm{reps}=5$) to reach chemical accuracy ($\sim 1 \textrm{mHa}$) over all bond distances.   The additional layer was chosen as layer 1 (dashed box) in circuit (\ref{vb:4site-ising}). By choosing the adiabatic optimization strategy over the default optimization, $\textrm{reps}=4$ (adiabatic) was also able to reach chemical accuracy for the full range of bond distances. 

When transferring the full cluster circuit, it is important to have a clear understanding of the qubit mapping: which qubit is mapped to which orbital.  In the transverse field Ising model, each site couples only to its neighbour sites.  This is in contrast with molecules, where qubits are associated with Hartree-Fock molecular orbitals.  The arrangement of the orbitals/qubits is in block order, with the first half encoding the information for the spin-up and the second half for the spin down. Therefore, the cluster circuit only makes the connection between spin-up and spin-down qubits in the second layer. The first layer of the circuit offers a similar situation, however, for \ch{H2} $q_0$ and $q_1$ are both spin-up orbitals, but the first qubit represents the $\sigma$ bonding and the second qubit represents the $\sigma^\ast$ antibonding orbital.  Therefore, it is important to consider the bonding/antibonding and spin-up/down arrangement when transferring the qubit mappings.
\subsection{6-qubit systems}
The procedure was repeated for several 6-qubit systems, including (1) the \ch{H2} molecule in the 6-31G basis set with PT mapping, (2) \ch{H4} and \ch{(H2)2} in the STO-3G basis with PT mapping, and (3) \ch{LiH} in STO-3G with frozen core and Z2 symmetry.  This selected set of test systems was chosen to test the universality of the full 6-qubit cluster circuit, as well as to benchmark accuracy and circuit depth with previous studies\cite{zhang:2022}.  

\begin{figure}[!htb]
\centering
    \hrule
    \vspace{\baselineskip}    
    Cluster A (Kekul\'e)\\
    \hfill\includegraphics[width=0.3\linewidth]{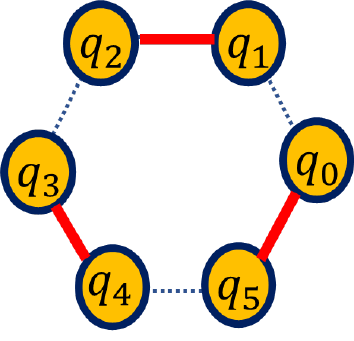}
    \hfill
    \includegraphics[width=0.3\linewidth]{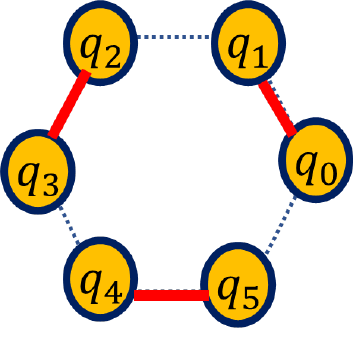}\hfill
    \vspace{\baselineskip}     
    \hrule
    \vspace{\baselineskip}
    Cluster B (Dewar)\\
     \hfill\includegraphics[width=0.3\linewidth]{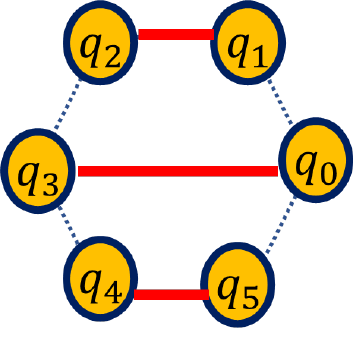}
  \hfill
    \includegraphics[width=0.3\linewidth]{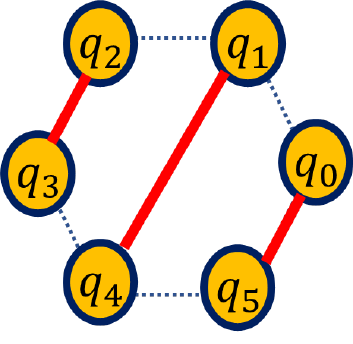}
    \hfill
    \includegraphics[width=0.3\linewidth]{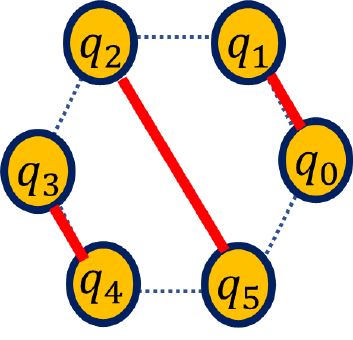}
    \hfill
     \vspace{\baselineskip}     
    \hrule   
     \vspace{\baselineskip}
    Cluster C (reduced-Dewar)\\
     \hfill\includegraphics[width=0.3\linewidth]{fig06c}
  \hfill
    \includegraphics[width=0.3\linewidth]{fig06d}\hfill
     \vspace{\baselineskip}     
    \hrule
  \caption{\label{fig6-6sitervb} Selected RVB structures of the 6-site ($q_i,i=0\dots5)$ transverse-field Ising model. Each pair of qubits connected with a red full line represents a cluster circuit unit (\ref{vb:cluster}).  Cluster A depicts a resonance which is commonly known as the Kekul\'e resonance, whereas Cluster B and C denote the regular and reduced Dewar structure respectively.}
\end{figure}

\begin{figure}[!htb]
\centering
\includegraphics{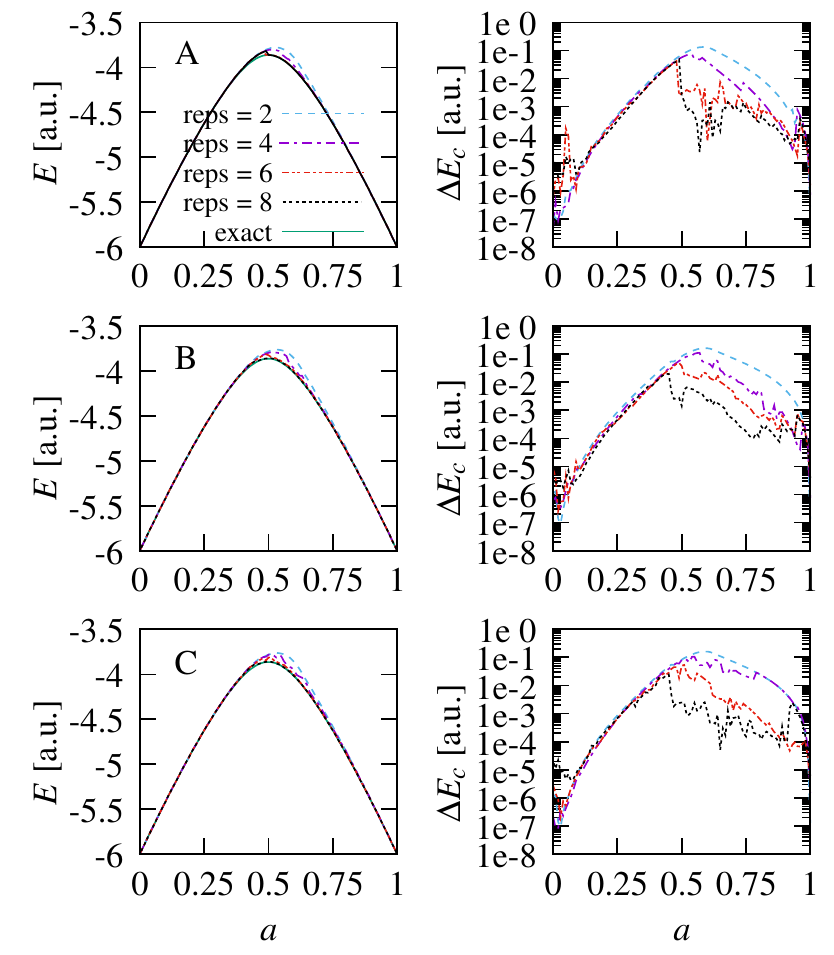}
\caption{\label{fig7-6siteising} Left: Comparison between the cluster circuit results with different $\textrm{reps}$ with the exact calculation for the 6-site Ising model. Right: missing correlation energy (\ref{4qubit:deltaE}) for corresponding reps. }
\end{figure}

First, the full cluster circuit was designed to accurately simulate the 6-site Ising model.  There are several possibilities to design the VB cluster circuits, corresponding to different unique pairings between the qubits.  This is a combinatorics problem in which one has to choose from $\frac{(2w)!}{2^w w!}$ unique pairings for a $2w$-qubit system, so it is important to select only the significant pairings.  The most common resonance structures associated with a cyclic hexagonal, such as benzene, are given in Figure \ref{fig6-6sitervb}, which are commonly referred to as the Kekul\'e (cluster A) and Dewar structures (cluster B) \cite{cooper:2002}.
For the purpose of this study, we also include a reduced Dewar structure (cluster C), in which the third configuration of the regular Dewar cluster (cluster B) has been removed.  The advantage of the reduced Dewar (C) over the regular Dewar (B) is that the former generates a $\textrm{reps}=2$ whereas the latter requires a $\textrm{reps}=3$ to describe a full resonance.  The definition of a reps is chosen consistently with the 4-qubit case, in which each layer of rotation gates followed by the \texttt{CNOT} entangles adds one unit to the total number of reps.  As we will see, the differences in accuracy for both cluster types are little and subtle.  As an example, the full cluster circuit for cluster C with $\textrm{reps}=2$ is given by
\scriptsize
\begin{equation}
\begin{array}{c}
\Qcircuit @C=1em @R=.7em {
\lstick{\ket{0}}  &\gate{R_Y(\theta_0)} & \ctrl{5} &\qw & \qw & \gate{R_Y(\theta_6)} & \ctrl{3} & \qw & \qw & \gate{R_Y(\theta_{12})}& \qw\\
\lstick{\ket{0}}  &\gate{R_Y(\theta_1)} & \qw &\ctrl{3} & \qw &\gate{R_Y(\theta_7)} & \qw & \ctrl{1} & \qw& \gate{R_Y(\theta_{13})}& \qw\\
\lstick{\ket{0}}  &\gate{R_Y(\theta_2)} & \qw &\qw & \ctrl{1} &\gate{R_Y(\theta_8)} & \qw & \targ & \qw& \gate{R_Y(\theta_{14})}& \qw\\
\lstick{\ket{0}}  &\gate{R_Y(\theta_3)} & \qw &\qw & \targ &\gate{R_Y(\theta_9)} & \targ & \qw  & \qw& \gate{R_Y(\theta_{15})}& \qw\\
\lstick{\ket{0}}  &\gate{R_Y(\theta_4)} & \qw &\targ & \qw &\gate{R_Y(\theta_{10})} & \ctrl{1} & \qw & \qw& \gate{R_Y(\theta_{16})}& \qw\\
\lstick{\ket{0}}  &\gate{R_Y(\theta_5)} & \targ &\qw & \qw &\gate{R_Y(\theta_{11})} & \targ & \qw  & \qw& \gate{R_Y(\theta_{17})}& \qw
\gategroup{1}{2}{6}{5}{0.8em}{--} \gategroup{1}{6}{6}{8}{0.8em}{--}
}
\end{array}\label{vb:6site-ising}
\end{equation}
\normalsize
with both RVB configurations (in dashed blocks) again depicted in Figure \ref{fig6-6sitervb}C.  Adding the last resonance of cluster B to the circuit would lead to the $\textrm{reps}=3$ of the regular Dewar cluster B, whereas repeating the first resonance of cluster B (or C) would give the $\textrm{reps}=3$ of the reduced-Dewar cluster C.  When increasing the reps in each cluster design, we add the resonances one by one from left to right as depicted in Figure \ref{fig6-6sitervb}.  
The VQE energy of the three cluster circuits with different reps for the 6-site Ising model are presented in Figure \ref{fig7-6siteising}.
Interestingly, all three cluster types (A, B, \& C) produce comparable results in terms of accuracy and reps convergence, with the Kekul\'e structure (Figure \ref{fig7-6siteising}A) displaying a slight advantage over both Dewar structures (Figure \ref{fig7-6siteising}B \& C), as can be expected for cyclic hexagonal systems, in which the Kekul\'e resonances are the dominant structures.  This can most likely be understood by the design of the full circuit, which is a concatenation of individual cluster units, rather than a pure linear combination of resonances.  This means that the correct information between, for instance, the Kekul\'e qubit pairing can be effectively realized by the Dewar structures, as long as the number of reps is sufficiently high.  This process of spreading information between sites via a concatenation of entanglers is well understood in the context of tensor network state approaches \cite{orus:2019}.  Furthermore, the small difference between the regular and reduced Dewar structure points towards a similar mechanism, in which the $\textrm{reps}=4$ of the regular Dewar (B) is more accurate than the reduced Dewar (C) result, however the difference largely disappears moving to larger reps.  This sharing of information process is corroborated by the observation that the accuracy of the circuit is not significantly improved when multiples of the intrinsic reps are reached in the cluster choices in Figure \ref{fig6-6sitervb}, for instance $\textrm{reps}=2n$ for the Kekul\'e (A) and reduced-Dewar cluster (C), and  $\textrm{reps}=3n$ for the Dewar cluster B.  This suggests that the passing of information between the qubits is effectively more important than the precise implementation of equally weighted resonances. 

So, we find that the correct choice of resonance pairing affects the accuracy of the VQE energy, however mostly with respect to the number of reps.  This is important for convergence purposes in the optimizer, as an increase in reps also induces an increase in variational parameters that need to be optimized, which is traditionally prone to issues with local minima and barren plateaus.  Again, we employed the SLSQP library in the Qiskit\cite{qiskit:2023} environment as a numerical optimizer.  Furthermore, due to the nature of the cluster unit (\ref{vb:mean field}), the variational space of a lower reps is not necessarily a subset of those of a higher reps.  This is a direct consequence of the \texttt{CNOT} entangler, which does not allow for the identity operator to be realised in each individual cluster unit (\ref{vb:mean field}).  Consequently, a set of optimized variational parameters for a certain reps is not necessarily a good seed for the optimization of higher reps.  In future work, we will investigate cluster units that allow for a better Lie group theoretical formulation \cite{devos:2018}, allowing for a better control over the numerical optimization however at the cost of introducing additional \texttt{CNOT} gates and increased circuit depth.  

In the following examples, we will present results for clusters (cluster A, B or C) that align best with chemical insight in terms of resonance structures, however results vary only slightly between the different options.  
\subsubsection{\ch{H2} in 6-31G basis set} 

\begin{figure}[h]
\centering
\includegraphics{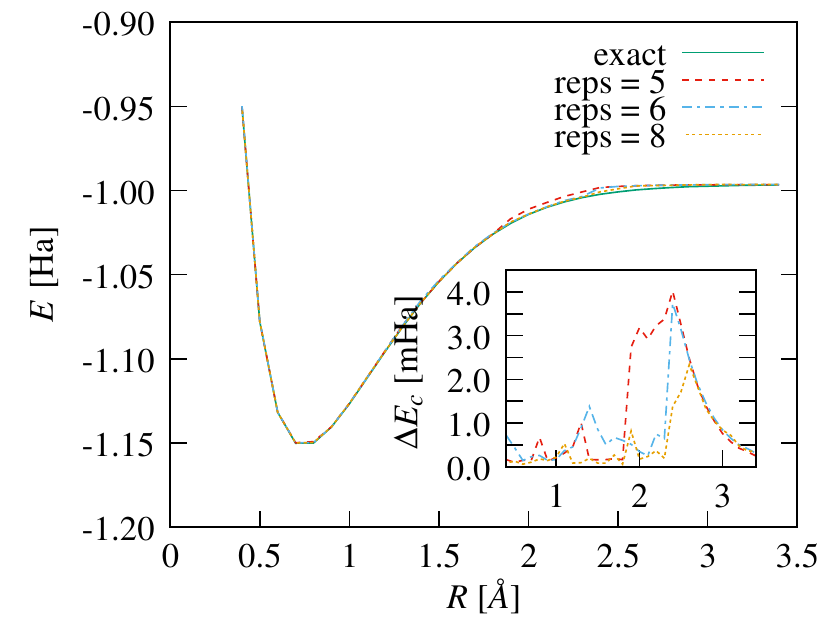}
\caption{\label{fig8-h2-631g}Energy prediction of the Dewar cluster (B) circuit for \ch{H2} in 6-31G basis using PT mapping for $\textrm{reps}=5$ (red dashed line), $\textrm{reps}=6$ (blue dot-dashed line) and $\textrm{reps}=8$ (yellow dotted line).  The exact reference energy is given by the green full line.  The missing correlation energy (\ref{4qubit:deltaE}) of all reps is given in the inset.}
\end{figure}

The PT mapping for \ch{H2} in a 6-31G basis set leads to a 6-qubit ($q_i,i=0\dots5$) circuit for this molecular system, three for the spin-up ($i=0,1,2$) and spin-down ($i=3,4,5$) orbitals each.  The 6-31G basis set introduces extra dynamical correlation with respect to the minimal basis sets, however, the bonding/antibonding resonance among the valence configurations of both hydrogens will remain the dominant contributions along the bond stretch.  In the PT mapping, this resonance is realized in a pairing between the $q_0$ and $q_3$ qubit.  Therefore, the Dewar structures (clusters B \& C) present themselves as best options based on chemical input.   Results for the Dewar cluster B for different reps are presented in Figure \ref{fig8-h2-631g}.  
Again, we observe that the VQE energy improves as the number of layers increases, showing that the cluster circuits are effectively able to capture dynamical correlation on top of static correlation.  

\subsubsection{\ch{H4} and \ch{(H2)2} in STO-3G basis set} 

\begin{figure}[h]
\centering
\includegraphics[width=0.4\textwidth]{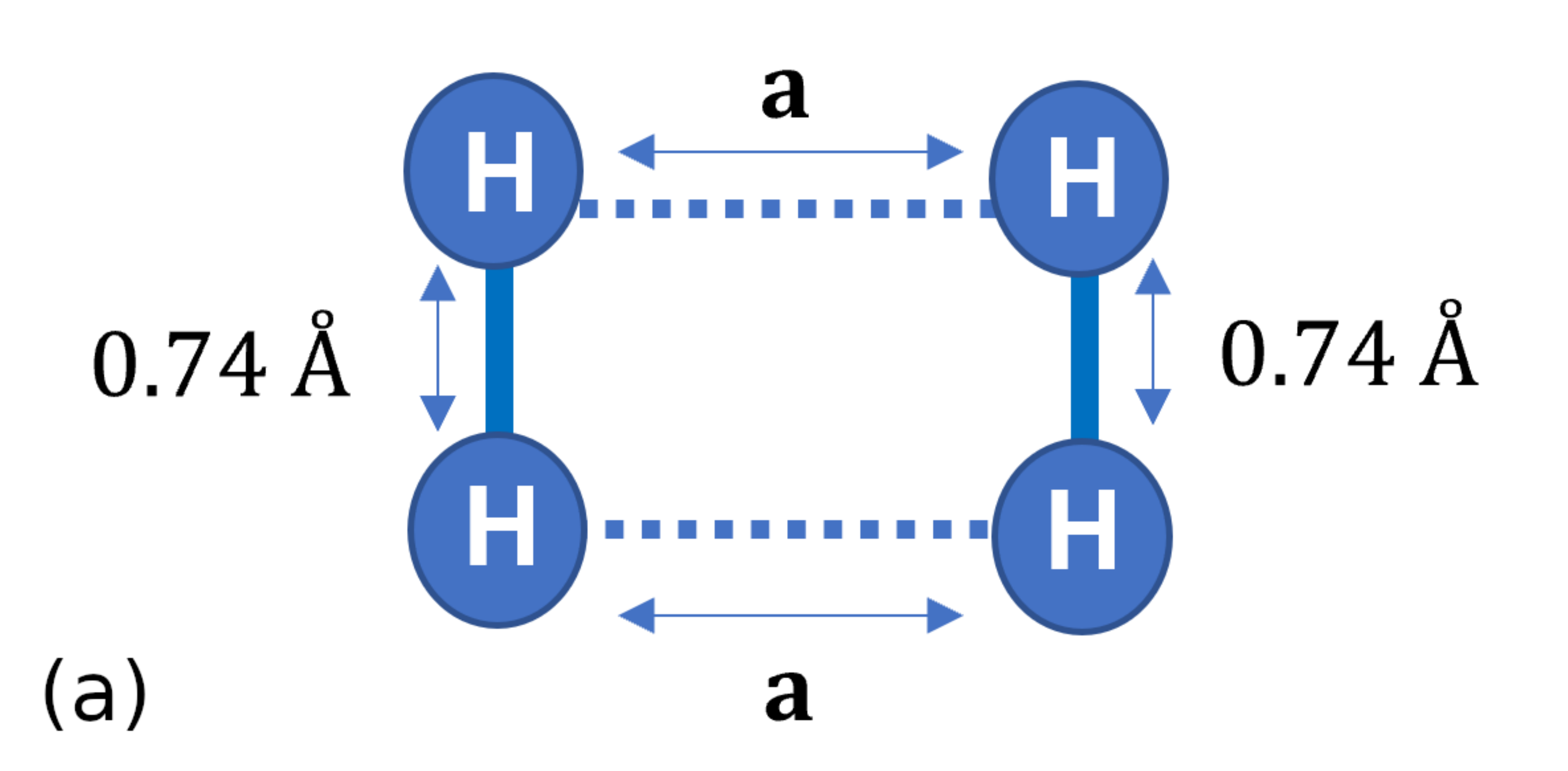}\\
\includegraphics[width=0.4\textwidth]{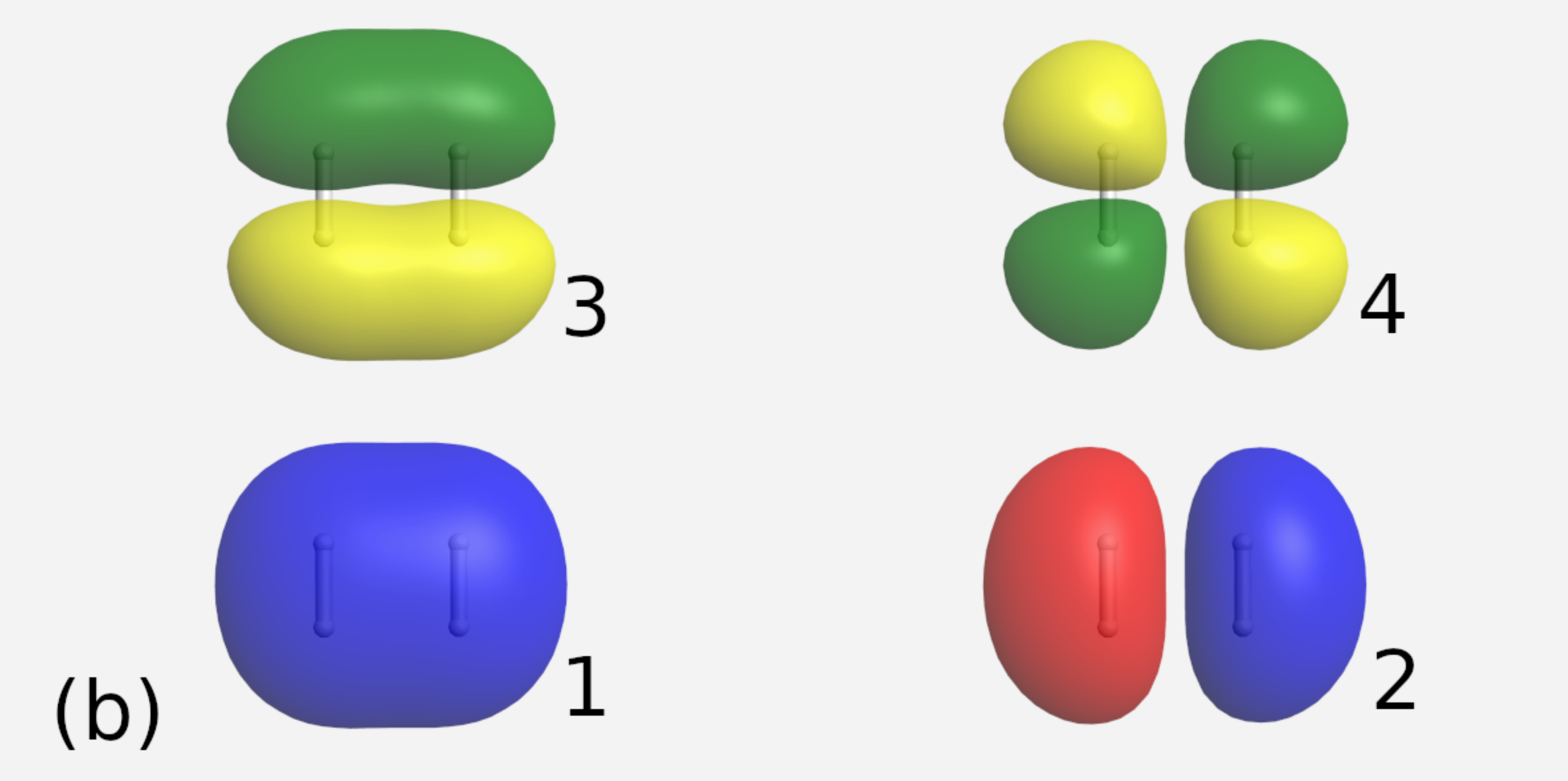}
\caption{(a) Geometry of \ch{H4}, in which the interatomic distance within both \ch{H2} molecules is fixed at $0.74$\AA, and the distance $a$ between the two moieties is varied. (b) Canonical RHF orbitals for \ch{H4} with $a=1.20\textrm{\AA}$.}
\label{fig9-h4}
\end{figure}

\begin{figure}[h]
\centering
\includegraphics{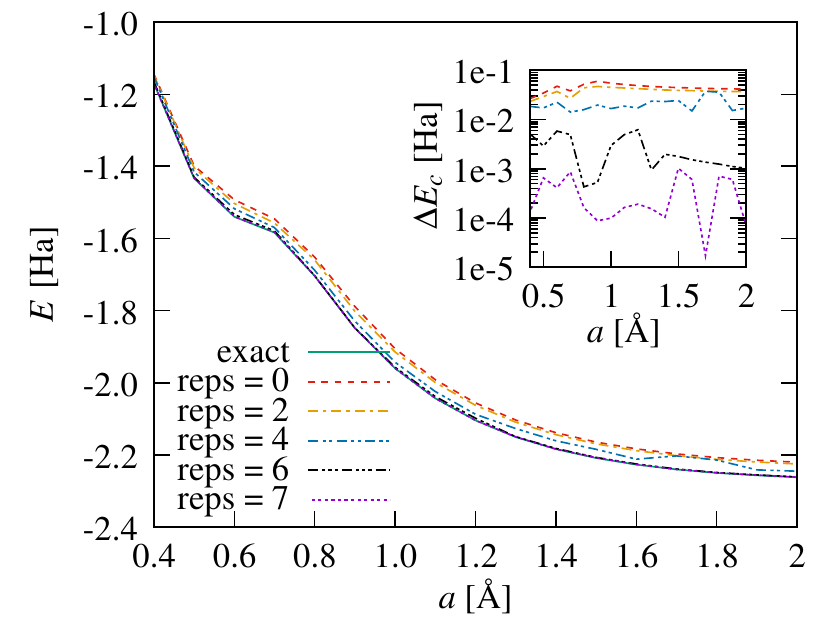}
\caption{\label{fig10-h4} Energy prediction of the reduced Dewar cluster circuit (C) for the \ch{H4} in STO-3G basis using PT mapping for various reps.  The inset shows the missing correlation energy (\ref{4qubit:deltaE}) for each reps.}
\end{figure}

The \ch{H4} model is often used as a first test case for strong correlation \cite{paldus:1993,limacher:2014}.  We investigate two different geometries.  The first geometry is the rectangular configuration in which the bond distance between the hydrogen atoms for the two \ch{H2} molecules are held fixed at 0.74\AA, and we let the two \ch{H2} approach (see Figure \ref{fig9-h4}a). 
The square configuration, for which $a=0.74\textrm{\AA}$ is a standard test for strong correlation.  The canonical orbitals of the rectangular configuration with $a=1.20\textrm{\AA}>0.74\textrm{\AA}$ are depicted in Figure \ref{fig9-h4}b, with the lower orbitals (1 \& 2) the occupied and the upper orbitals (3 \& 4) the virtuals in the restricted Hartree-Fock (RHF) configuration.  The exact wave function of the rectangular \ch{H4} structure is a resonance between a pair of electrons between orbitals 2 and 3, which corresponds to a pairing between qubits $q_1$ and $q_4$ in the PT mapping.  Therefore, we select the reduced Dewar cluster C (see Figure \ref{fig6-6sitervb}) to design the full circuit.  The results are depicted in Figure \ref{fig10-h4},
in which we observe a gradual increase in accuracy as the number of reps is increased. 

\begin{figure}[h]
\centering
\includegraphics[width=0.4\textwidth]{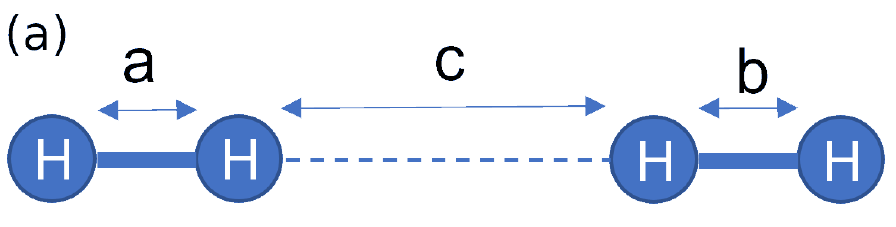}\\
\includegraphics[width=0.4\textwidth]{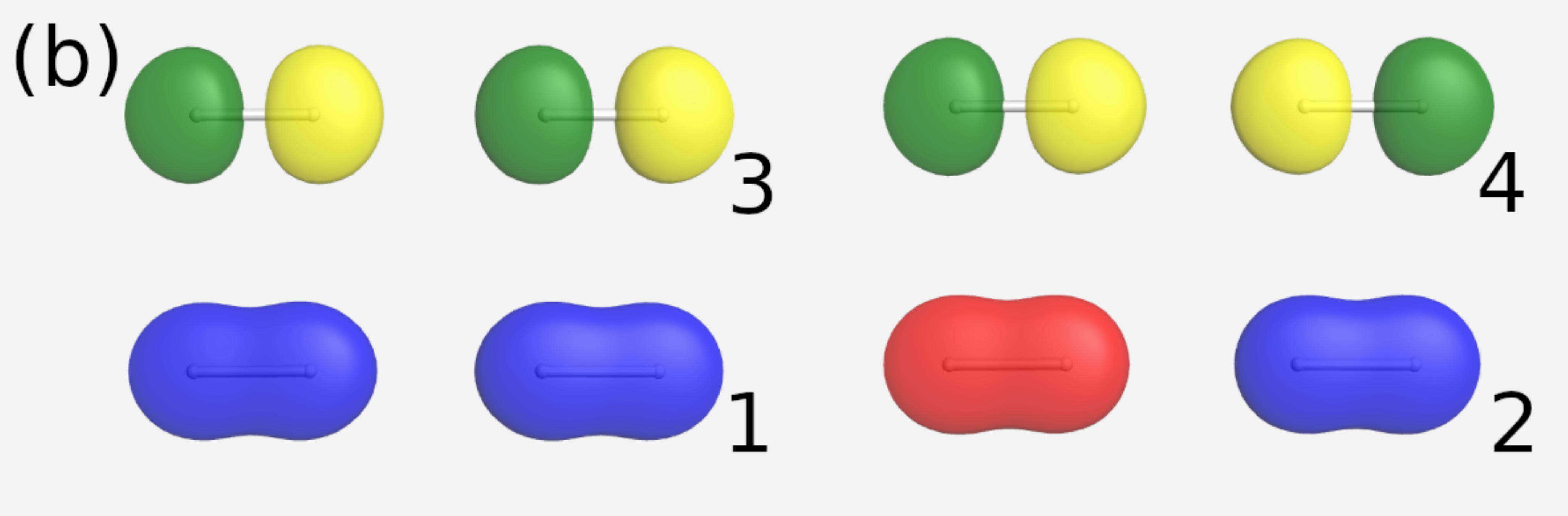}
\caption{(a) Geometry of \ch{(H2)2}, with the leftmost atom of each \ch{H2} fixed, and the relative distances given by $b=3.00\textrm{\AA}-a$ and $c=4.50\textrm{\AA}-a$. (b) Canonical RHF orbitals for \ch{(H2)2} with $a=1.50\textrm{\AA}$.}
\label{fig10-h22}
\end{figure}

\begin{figure}[h]
\centering
\includegraphics{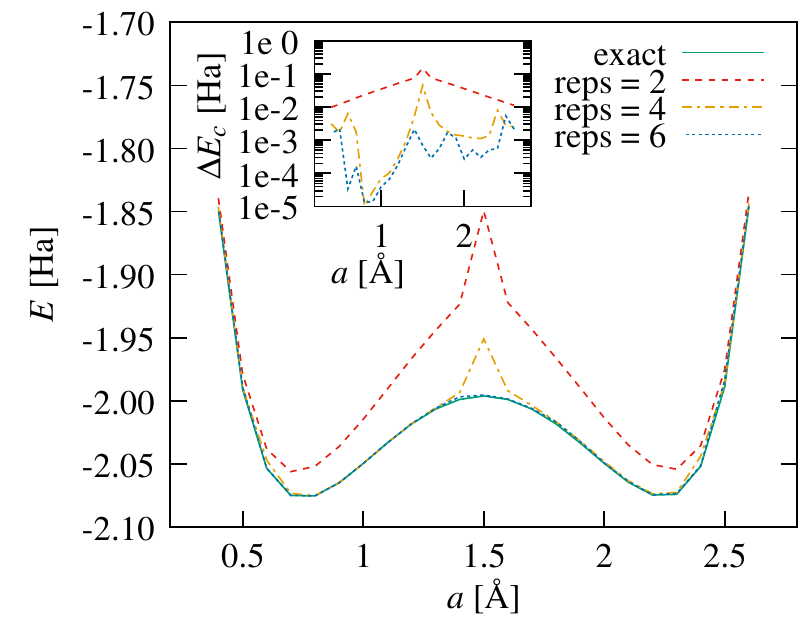}
\caption{\label{fig12-h22}Energy prediction of the regular Dewar cluster circuit (B) for \ch{(H2)2} for several reps. The inset depicts the missing correlation energy (\ref{4qubit:deltaE}) for each reps. }
\end{figure}

In the second geometry (see Figure \ref{fig10-h22}a),
the two \ch{H2} molecules are held in a linear configuration in which the position of the left hydrogen atom in each \ch{H2} is held fixed. Then we change the position of the right hydrogen atom along the axis between these two molecules according to the rule $a\in [0.50\AA,2.60\AA]$, $b=3.00\textrm{\AA}-a$ and $c=4.50\textrm{\AA}-a$.  The canonical RHF orbitals for $a=1.50\textrm{\AA}$ are depicted in Figure \ref{fig10-h22}b.  The linear system is not dominated by a single pair resonance, however many different configurations become equally important.  For this reason, we select the regular Dewar cluster (B) to design the full circuit as multiple of the resonances are present in this cluster circuit within the PT mapping.  The VQE results for several reps are depicted in Figure \ref{fig12-h22}, 
in which we notice that the exact result is obtained within chemical accuracy (mHa) at the reps = 6 level.
\subsubsection{\ch{LiH} in STO-3G basis set and Z2 symmetry} 

\begin{figure}[h]
\centering
\includegraphics{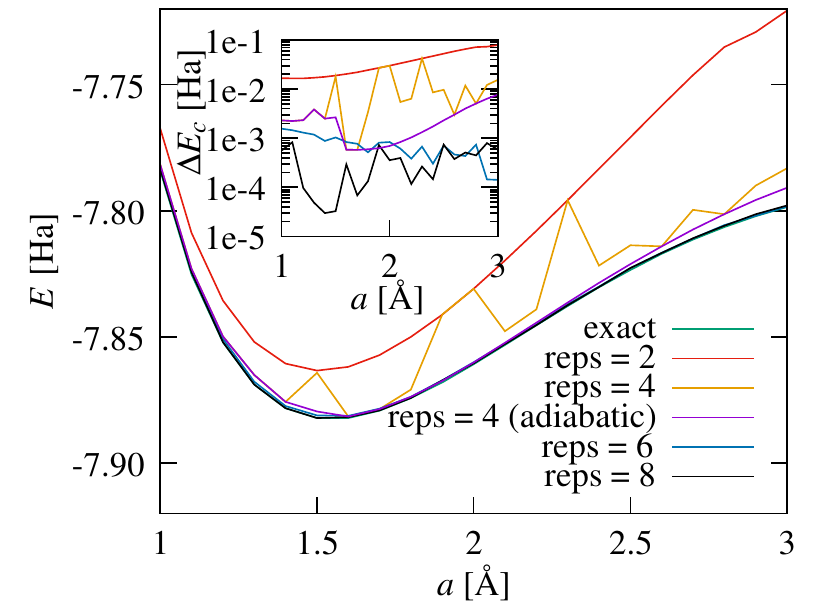}
\caption{\label{fig12-lih}Energy prediction of the reduced Dewar cluster circuit (C) for \ch{LiH} in STO-3G basis using PT mapping for various reps and optimization strategies.  The inset depicts the missing correlation energy (\ref{4qubit:deltaE}) for each reps.}
\end{figure}

We employed a reduced Dewar cluster C with $\textrm{reps}=4, 6, 8$ for the LiH molecule in STO-3G basis with Z2 symmetry, for which the results are depicted in Figure \ref{fig12-lih}.
We notice an improvement of the energy with increasing reps, with chemical accuracy obtained at the $\textrm{reps}=8$ level.  The $\textrm{reps}=4$ results reveal the existence of local minima in the optimization process, however these are apparently overcome with increasing reps.  The adiabatic optimization strategy  remedies the occurrence of the highly pronounced local minima at the $\textrm{reps}=4$ (adiabatic) level, however more reps are required to reach chemical accuracy in the bond-breaking regime. 

In Table \ref{table.1}, we present the missing correlation energy $\Delta E_c$ (\ref{4qubit:deltaE}) for the 6-qubit systems presented.  For all systems, a large portion of the missing correlation for the equilibrium distances for the simulated molecules can be captured by the cluster circuit with the $\textrm{reps}=6$ (and corresponding depth of 13). For the Ising model, however, we must increase the reps to $\textrm{reps}=8$ in order to reduce the discrepancy between the results of the exact and cluster circuits.

\begin{table}[h!]
\centering
\caption{Circuit depth and the corresponding missing correlation energy ($\Delta E_{c}$) for each presented 6-qubit system.}
\begin{tabular}[t]{lcrrr}
  \hline
   System & cluster & reps & depth & $\Delta E_{c}$ \\
  \hline\hline
  6-site IM       & A & 2 &  5 & 0.07028\\
  ($a=0.5$)       &   & 4 &  9 & 0.05662\\  
                  &   & 6 & 13 & 0.00296\\
                  &   & 8 & 17 & 0.00250\\
  6-site IM       & B & 2 & 5 & 0.08257\\
  ($a=0.5$       &   & 4 & 9 & 0.06482\\ 
                  &   & 6 & 13 & 0.03775\\
                  &   & 8 & 17 & 0.00142\\
  6-site IM       & C & 2 & 5 & 0.08257\\
  ($a=0.5$       &   & 4 & 9 & 0.08128\\
                  &   & 6 & 13 & 0.05027\\   
                  &   & 8 & 17 & 0.00245\\                                
                  \hline
  \ch{H2} (6-31G) & B & 5 & 11 & 0.00014\\
  ($R=0.70$\AA)    &   & 6 & 13 & 0.00023\\
                  &   & 8 & 17 & 0.00010\\
                  \hline
  \ch{H4}         & C & 0 & 1 & 0.04057\\
  ($a=0.74$\AA)   &   & 2 & 5 & 0.02287\\
                  &   & 4 & 9 & 0.01487\\  
                  &   & 6 & 13 & 0.00002\\
                  &   & 7 & 15 & 0.00001\\                    
  \hline                  
  \ch{(H2)2}      & B & 2 & 5 & 0.14665\\
  ($a=1.50$\AA)    &   & 4 & 9 & 0.04528\\  
                  &   & 6 & 13 & 0.00070\\  
  \hline
  LiH             & C & 0 &  1 & 0.53875\\
  ($R=1.55$\AA)  &   & 2 &  5 & 0.01942\\
                  &   & 4 &  9 & 0.00268\\                  
                  &   & 6 & 13 & 0.00089\\
                  &   & 8 & 17 & 0.00003\\
  \hline
\end{tabular}
\label{table.1}
\end{table}
For the purpose of comparison, it should be noted that the circuit depths of the VQE with the 1 layer Qubit Unitary Coupled-Cluster Singles and Doubles (QUCCSD) ansatz are 352 for LiH \cite{zhang:2022} and the number of CNOT gates using the ADAPT-VQE approach\cite{tang:2021} is of the order of $10^3$, which is significantly higher than required for the modular cluster circuits in the present work.  

We also compared our circuit parameters for the equilibrium distance of the LiH molecule to ADAPT-VQE, ClusterVQE and iQCC, as reported in ref \cite{zhang:2022}.  We confront the circuit depth, number of variational parameters (iterations) at chemical accuracy in Table \ref{table.3}).  
\begin{table}[h!]
\centering
\caption{Circuit depth, number of variational parameters (iterations) and number of Pauli words of ADAPT-VQE, ClusterVQE, and iQCC techniques with the QUCCSD ansatz for LiH molecule at equilibrium distance in comparison to the cluster circuit introduced in this work.  Numbers taken from \cite{zhang:2022}.}
\begin{tabular}[!htb]{lrrr}
  \hline
  method & depth & iterations & Pauli words\\
  \hline\hline
  ADAPT-VQE  & 103 &  6 & 275\\
  ClusterVQE &  69 &  6 & $\sim500$\\
  iQCC       &  12 &  6 & $\sim1000$\\
  This work  &  13 & 36 & 230\\
  \hline
\end{tabular}
\label{table.3}
\end{table}
All three methods reach chemical accuracy at 6 iterations, which is lower than the $36$ parameters that are introduced in the cluster circuit.  On the flip side, the circuit depth of our circuit circuits is considerably lower, except for the iQCC method.  However, it should be noted that the latter requires more Pauli words evaluations, whereas the number of Pauli words of the cluster circuit is slightly lower than ADAPT-VQE, mostly due to the extra incorporation of the Z2 symmetry in our simulations.  The ClusterVQE approach shows a trade off between the circuit depth and number of Pauli words that need to be measured.  For stretched bond configurations ($R=2.40$\AA), ADAPT-VQE and ClusterVQE require more iterations (up to 18) to reach chemical accuracy, however the number of Pauli words to be measured remains unaltered.  In contrast, the parameters reported in Table \ref{table.3} are constant for all geometries in this work.  iQCC was not able to converge to chemical accuracy.  

\section{Conclusion}
\label{sec:4}

We introduced the cluster circuit as a modular unit to construct full quantum circuit for the VQE with low circuit depth.  The design of the cluster circuit allows for a chemical interpretation in terms of valence bond structures via the concatenation of rotation gates and entanglers.  The choice of particular cluster depends on the particular fermion-qubit mapping employed, which is preferably local, as one needs a clear picture of the orbitals involved in the valence bond resonances.  We tested the approach for two, four and six qubit systems.  An interesting observation is that the required circuit depth is approximately double the number of qubits in all cases in order to reach full correspondence with the exact result.  As the number of variational parameters scales linearly with the number of reps, this introduces a significant variational space.  One can say that the variational density per circuit depth is high in the cluster circuits.  It should be noted however that increasing the reps consistently improves accuracy over all bond lengths, pointing out that different aspects of the circuit may be active in different regimes.  Moving forward, it will be interesting to investigate whether the empirically observed linear relation between circuit depth and circuit width still holds, and whether the mapping of the simple cluster circuits can be further optimized, either via their Lie algebraic properties, through further chemical input or (symmetry) constraints.  In the present paper, a fixed template was used for all bond distances, from equilibrium to full dissociation, despite the different nature in correlations (dynamic vs static).  It is reasonable to assume that an adaptive template can introduce chemically motivated simplifications in the circuit.  We will address this in future work.

\begin{acknowledgement}

SDB and SEG acknowledge the Canada Research Chair program, the CFI, NSERC Discovery Grant program and NBIF for financial support.  We are grateful to Dr.\ Jakob Kottmann for insightful discussions about Valence Bond theory and quantum computing circuit design.  The anonymous reviewers are acknowledged for their insightful questions and comments. 

\end{acknowledgement}


\section{Data Availability Statement}
A Jupyter Notebook of the simulations can be found at\\
 \texttt{https://github.com/QuNB-Repo/QCChem}.

%
%
%

\bibliography{ghasempouri-clustercircuits-rerevision}

\end{document}